\documentclass[11pt,twoside]{article}
\usepackage{asp2010}

\resetcounters

\begin{document}

\title{Rapid optical variations in KT Eri}
\author{K. I\l{}kiewicz, E. \'{S}wierczy\'{n}ski, C. Ga\l{}an, M. Cika\l{}a and T. Tomov
\affil{Centre for Astronomy, Faculty of Physics, Astronomy and Informatics, Nicolaus Copernicus University, Grudziadzka 5, 87-100 Torun}
}

\begin{abstract}
We present optical photometric monitoring of KT Eri (Nova Eridani 2009), a He/N very fast nova which oubursted in November 2009. Our observations include $BVR_cI_c$ brightness estimations as well as monitoring of the rapid brightness variations in V band. The characteristic times of these rapid changes are studied and compared with the observed in other novae.
\end{abstract}

\section{Introduction}
KT Eri was discovered by \citet{2009CBET.2050....1I} in November 2009 at high galactic latitude. Due to this unusual position it is the first classical nova observed in the constellation Eridanus. \citet{2009ATel.2327....1R} estimated that the maximum brightness $V\sim5.\!\!^\mathrm{m}6$ occurred on November 15 which shows that the nova is very fast with $t_\mathrm{2}\sim8$\,days and $t_\mathrm{3}\sim15$\,days. \citet{2009CBET.2053....3M} classified KT Eri as He/N nova. The pre-outburst light curve shows large variations in the optical and an average brightness $\sim 15^\mathrm{m}$ \citep{2009ATel.2331....1D}. Because of the relatively low amplitude ($\sim 10^\mathrm{m}$), unusual for very fast classical novae, it was suggested \citep{2010ApJ...724..480H} that KT Eri is in fact recurrent nova, but no evidence of other outbursts are yet available. Studying the historical light curve of the nova, \citet{2012A&A...537A..34J} found a possible orbital period of 737 days and suggested that the secondary component could be a RGB star.

\hspace{1em} Quasi-periodic Oscillations (QPOs) are rapid brightness variations often observed in cataclysmic variables. Their typical periods are from 3s to $\sim$1000s usually unstable and with low amplitudes. QPOs with short periods are more profoundly studied, although all QPOs nature is uncertain (see \citeauthor{2004PASP..116..115W} \citeyear{2004PASP..116..115W}, \citeauthor{2008AIPC.1054..101W} \citeyear{2008AIPC.1054..101W}).

\hspace{1em} \citet{2010ATel.2392....1B} were the first to report on possible rapid brightness variations in optical observations of  KT Eri obtained from day 17 to day 65 after the maximum. Unfortunately, the frequency of their data sampling was too small for periodogram analysis. Very short $P\sim35.09$ sec QPO were observed by \citet{2010ATel.2423....1B} in X-ray between 67 and 79 days after the maximum. Long-term variability of KT~Eri with a period of 56.7 days was reported by \citet{2011ASPC..451..271H}, interpreted by them as rotation of the hot spot on the accretion disk.

\section{Observations}
We performed $BVR_cI_c$ photometry in ten nights between 26 November 2009 and 25 October 2012. During six of these nights time-resolved monitoring in $V$ was carried out as well. Two identical 60 cm Cassegrain telescopes in Torun Observatory (Poland) and Rozhen Observatory (Bulgaria) were used. Most of the observations were carried out at the Torun Observatory by the use of a SBIG STL-1001E CCD camera. In one occasion we obtained observations in Rozhen Observatory using a PL 9000 CCD-camera.

\hspace{1em} The standard IRAF procedures were used to reduce the CCD observations and to perform aperture photometry. For comparison we used the stars numbered 128, 134 and 142 in the AAVSO sequence for KT Eri. The accuracy of our measurements in all filters is better than $0.\!\!^\mathrm{m}03$. The estimated $BVR_cI_c$ magnitudes of KT Eri are shown in Table~\ref{tab1}. The data for the time-resolved photometry is presented in Table~\ref{tab2}.
\begin{table}[h!]
 \centering
 \caption{Our KT Eri $BVR_cI_c$ magnitudes.}\label{tab1}
 \vspace*{1ex}
\begin{tabular}{ | c | c | c | c | c | c | c | c | c | }\hline
HJD		& Observatory &	B	& V	& R 	& I	\\\hline
2455162.50	& Torun &	 8.52	& 8.37	 &7.61	& 7.60	\\
2455168.48	& Torun & 	 8.81	& 8.79	 &8.11	& 8.32	\\
2455210.38	& Torun &	10.35	&10.07	&10.01	& 9.88	\\
2455231.24	& Torun &	11.58	&11.24	&11.17	& 11.21\\
2455235.24	& Torun &	11.64	&11.52	&11.30	& 11.29\\
2455236.26	& Torun &	11.83	&11.63	&11.52	& 11.42\\
2455237.28	& Torun &	11.63	&11.31  &11.20 & 11.18 	\\
2455267.23	& Torun &	11.85  &11.55  &11.57 & 11.33 	\\
2455592.38	& Torun &	14.30  &13.98  &13.96 & 13.59 	\\
2456226.42	& Rozhen &	15.45	&15.03	&15.05	& 	\\
\hline
\end{tabular}
\end{table}
\section{Results}
Fourier analysis of our time-resolved observations was performed with the~\textit{Period04} program \citep{2005CoAst.146...53L}. To estimate the errors of the obtained periods we used \citet{1999DSSN...13...28M} method. The analysis revealed QPOs in three of the nights while we did not succeed to find any quasi periodicity in the other three nights.

\hspace{1em} The derived QPO periods together with the amplitude of the fitted sine functions and the accuracy of the fits are shown in Table~\ref{tab2}. The large scattering around the fitted sines is obvious in all the cases. The estimated values for the QPO periods are affected by relatively big errors as well. Of particular interest is the QPOs period behaviour during the $\sim$4.5 hours long monitoring 133 days after the maximum. The period is apparently different during the first and the second halves of the observations, increasing from 3809 sec to 6560 sec (Fig.~1). Most uncertain, shortest period 3530 sec and smallest amplitude $0.\!\!^\mathrm{m}005$ were observed on day 142. 1135 days after the maximum when the nova was close to its pre-outburst brightness we detected the longest period of 10440 sec with an amplitude about $0.\!\!^\mathrm{m}04$. Our time-resolved observations on this date cover 1.7 cycles suggesting that such long period could be plausible.
\begin{figure}[!h]
\label{fig1}
\centering
    \includegraphics[width=0.9\textwidth]{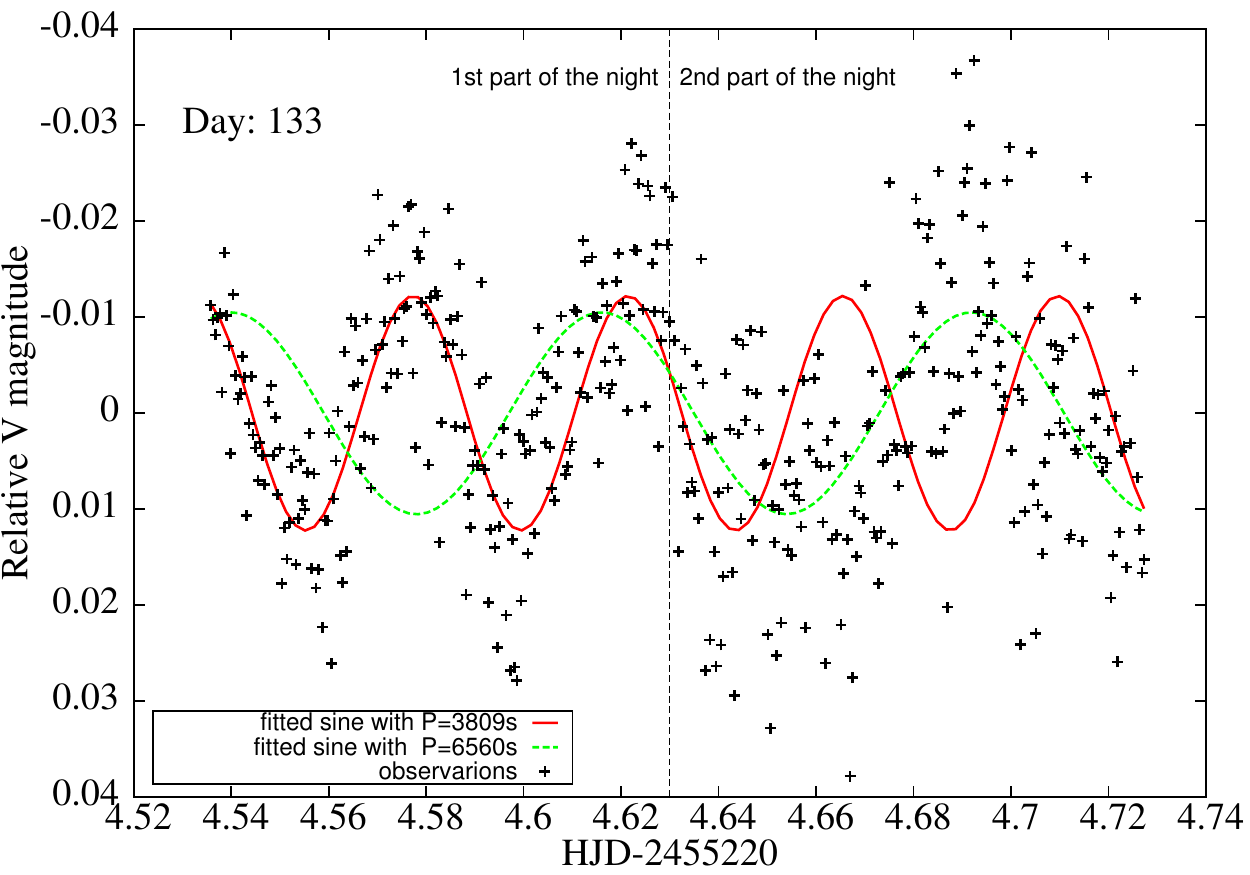}
\caption{The time-resolved photometry in $V$ filter 133 days after the maximum, fitted with two sine functions, is presented. The sine function shown as a solid line is fitted with P=3809 sec, estimated from the 1st half of the data, and the sine function shown as a dashed line is fitted with P=6560 sec, estimated from the 2nd half of the data.}
\end{figure}

\begin{table}[h!]
 \centering
 \caption{KT Eri $V$ band time-resolved photometry. The observations on the 1135 day after maximum were carried out at the Rozhen observatory. The rest of the observations were carried out at the Torun observatory.}\label{tab2}
 \vspace*{1ex}
\begin{tabular}{ | c | c | c | c | c |}\hline
Days &  Integration time & Duration  & QPO period & Amplitude \\
since maximum & [sec] & [sec] &  [sec] & mag\\\hline
119 & 90 & 9483  & not detected & \\
133 & 60 & 16546  & 3809$\pm$62, & 0.011$\pm$0.001, \\
 &       &        & 6560$\pm$260 & 0.012$\pm$0.001\\
142 & 30 & 6921  & 3530$\pm$210 & 0.005$\pm$0.008\\
145 & 30 & 12402  & not detected & \\
146 & 60 & 9683  & not detected & \\
1135 & 20 & 17750 & 10440$\pm$550 & 0.043$\pm$0.024\\
\hline
\end{tabular}
\end{table}
\section{Discussion}
The results of our time-resolved photometry show that the KT Eri post-outburst optical QPOs appear and disappear in several day intervals. Taking into account the accuracy of our observations we cannot be sure that the QPOs really are absent. It is possible that we are not able to detect the ones with lower amplitudes.

\hspace{1em} There are several novae which show a variety of QPO periods: IX Vel $\sim$500 sec, V533 Her $\sim$1400 sec, BT Mon $\sim$1800 sec, GK Per $\sim$5000 sec, V842 Cen $\sim$750-1300 sec (Warner 2004), V2468 Cyg $\sim$1260-3000 sec \citep{2012MmSAI..83..767C}. Most of these periods, excepting GK Per QPOs period are shorter in comparison to ones of KT Eri.

\hspace{1em}  It is interesting to note, that about two years after the outburst of the recurrent nova RS Oph in 2006,  \citet{2009ASPC..404...72V} found optical QPOs with a period $\sim$3000 sec. Very close to the observed in KT Eri on days 133 and 142.  Furthermore,  \citet{2011ApJ...727..124O} detected X-ray QPOs with a period $\sim$35 sec during the supersoft phase of RS Oph, similar to the reported by Beardmoreday et al. (2010) for KT Eri.

\hspace{1em} The resemblance of the KT Eri and RS Oph quasi periodic oscillations may be an additional argument in favor of the suggestion that KT Eri is in fact a recurrent nova. But the large variety of the periods and the low stability of the QPOs observed in novae make this argument insignificant.

\acknowledgements We greatly acknowledge the comparison stars magnitudes used in this research from the AAVSO International Database contributed by observers from Astrokolhoz Observatory. KI was in part supported by the local organizing committee of the \textit{Stella Novae: Future and Past Decades} conference.

\bibliographystyle{asp2010}
\bibliography{Ilkiewicz.bib}
\end{document}